\documentclass[prb,fleqn,12pt]{revtex4}
\usepackage{epsfig}
\usepackage{graphicx}
\usepackage{amsfonts}
\begin{document}
\title{Correlation effects on magnetic frustration \\ in the triangular-lattice Hubbard model}
\author{Saptarshi Ghosh}
\email{gsap@iitk.ac.in}
\author{Avinash Singh}
\affiliation{Department of Physics, Indian Institute of Technology Kanpur - 208016}
\begin{abstract}
Evolution of the magnetic response function in the triangular-lattice Hubbard model is studied with interaction strength within a systematic inverse-degeneracy expansion scheme which incorporates self-energy and vertex corrections and explicitly preserves the spin-rotation symmetry. It is shown that at half filling the response function goes through a nearly dispersionless regime around K for intermediate coupling strength, before undergoing an inversion at strong coupling, resulting in maximum response at the K point, consistent with the expected $120^\circ$ AF instability. Effects of finite hole/electron doping on the magnetic response function are also examined.
\end{abstract}
\maketitle
\newpage
\section{Introduction}
There has been renewed interest in correlated electron systems on triangular lattices, as evidenced by recent studies of antiferromagnetism, superconductivity and metal-insulator transition in the organic systems
$\rm \kappa -(BEDT-TTF)_2 X$,\cite{review1,review2} the discovery of superconductivity in $\rm Na_x Co O_2 . y H_2 O$,\cite{watersup} the observation of low-temperature insulating phases in some $\sqrt{3}$-adlayer structures such as K on Si[111],\cite{weitering} and quasi two-dimensional $120^\circ$ spin ordering and spin-wave excitations in $\rm Rb Fe (MoO_4)_2$ (Refs. 5,6) and the multiferroic materials $\rm Y Mn O_3$ and $\rm Ho Mn O_3$.\cite{sato,holmium1,holmium2,ymno3}

Transverse spin fluctuations in the $120^\circ$ ordered antiferromagnetic (AF) state of the triangular-lattice Hubbard model at half filling were recently investigated in the full range of interaction strength $U$.\cite{tri,triself} While a stable AF state was obtained in the strong-coupling limit, with identical spin-wave dispersion as for the equivalent QHAF,
with decreasing $U$ the spin stiffness was found to vanish at $U \approx 6$. The loss of magnetic order due to divergent quantum fluctuations yields a magnetic phase transition to a quantum spin-disordered insulator, which is relevant to the spin-liquid state and Mott transition in the organic systems $\rm \kappa -(BEDT-TTF)_2 X$. The vanishing spin stiffness implies that the triangular-lattice Hubbard model exhibits, besides the intrinsic geometrical frustration of the triangular lattice, an additional $U$-controlled frustration due to competing extended-range spin couplings generated at finite $U$. The existence of stable $120^\circ$ AF ordering at large $U$, but the vanishing spin stiffness (as well as energy $\omega_M$) indicates competition with other  magnetic orderings with decreasing $U$.

In view of this enhanced magnetic frustration in the triangular-lattice Hubbard model at finite $U$, it is therefore of interest to examine how the magnetic response function evolves with increasing interaction strength in the correlated paramagnetic state. Even at the bare level, the magnetic response function shows very rich behaviour [Fig. 1].  The comparable magnetic response at different symmetry points in the Brillouin zone, correponding to very different magnetic orderings, represents the weak-coupling picture of competing orders and magnetic frustration in the triangular-lattice paramagnet. Furthermore, the bare response is not maximum at the K point corresponding to $120^\circ$ AF ordering. In view of the expected instability towards the $120^\circ$ AF ordering at strong coupling, it will be desirable to develop an approach wherein the evolution of the magnetic response with increasing interaction strength $U$ is consistent with this AF instability.

Given that the $120^\circ$ AF state is stable in the strong-coupling limit, it would be desirable to use a many-body approach which continuously interpolates to the spontaneously-broken-symmetry AF state and yields a proper description of the Goldstone mode and spin waves by preserving the spin-rotation symmetry. From this viewpoint, evaluating many-body corrections reliably in the intermediate and strong-coupling regimes remains a challenge. Schemes such as the dynamical mean-field theory (DMFT) and fluctuation-exchange (FLEX) approximation, although providing powerful tools for studying the correlated paramagnet, do not continue into the broken-symmetry state without breaking the essential spin-rotation symmetry, while the two-particle self-consistent (TPSC) approximation does not the renormalize the momentum structure of the magnetic response function.

In this paper we will use a systematic many-body expansion scheme to investigate the momentum-dependent magnetic response function, and study its evolution with increasing interaction strength. For this purpose we will use a systematic inverse-degeneracy ($1/\cal N$) expansion scheme which explicitly preserves the spin-rotational symmetry by including self-energy diagrams as well as the corresponding vertex corrections. The importance of including vertex corrections in preserving spin-rotation symmetry has been highlighted in the context of paramagnon corrections\cite{ma} in He$^3$, the antiferromagnetic ground state,\cite{quantum} and the ferromagnetic ground state.\cite{hertz,singh_ferro} Indeed, we will show that the dominant quantum correction to the response function arises from the vertex corrections, signifying suppression due to particle-particle correlations. Therefore, from the paramagnetic side, vertex corrections play a dominant role in suppressing magnetic frustration and stabilizing the $120^\circ$ AF ordering at half filling.

The Hubbard model on a triangular lattice has been studied recently using a variety of tools. The zero-temperature phase diagram has been studied using the slave boson (SB) technique and the exact diagonalization.\cite{gazza_1994,slavebos} The mean-field SB approach yields a rich phase diagram qualitatively resembling the Hartree-Fock results.\cite{hf1,hf2} The non-magnetic insulating state near the Mott transition has been studied using the path integral renormalization group method,\cite{pirg} in which the HF results are systematically improved to reach the true ground state by taking account of quantum fluctuations. Results show a generic emergence of a non-magnetic insulating state sandwiched by a Mott metal-insulator transition and an AF transition. One-electron density of states has been examined using the quantum Monte Carlo method,\cite{qmc_green} showing a pseudogap development for intermediate $U$, accompanied by two peaks in the spin structure factor, signaling the formation of a spiral spin density wave (SDW). A weak-coupling RG analysis applied to the anisotropic triangular lattice shows that frustration suppresses the antiferromagnetic instability in favour of a superconducting instability.\cite{weakcoup} A magnetic field induced exotic spin-triplet superconductivity has been proposed
having strong ferromagnetic fluctuations.\cite{triplet} 

A spin-liquid type non-magnetic insulating (NMI) state sandwiched between a weak-coupling PM state and a strong-coupling AFI state has also been obtained for the $t-t'$-Hubbard model on a square lattice and an anisotropic triangular lattice using the path integral renormalization group method.\cite{pirg2,pirg3} The NMI state has been recently suggested to be a new type of degenerate quantum spin phase having gapless and dispersionless spin excitations.\cite{pirg3} At the same time, this result of an intervening NMI state is in contradiction to the earlier finding of an intermediate metallic antiferromagnetic state (AFM).\cite{duffy} In the context of $\rm \kappa -(BEDT-TTF)_2 Cu_2 (CN)_3 $, spin-liquid phases near the Mott transition in the Hubbard model have also been studied within the U(1) gauge theory.\cite{lee_2005} 

The organization of the paper is as follows. The inverse-degeneracy expansion scheme is briefly reviewed in section II.
The order-$1/{\cal N}$ diagrams for the irreducible propagator and their expressions are given in section III. Results at half filling and for finite electron/hole doping are discussed in sections IV and V, and conclusions are presented in section VI. Evaluation of the fermion vertices by integrating out the fermion energy-momentum degrees of freedom are illustrated in Appendix A. Emergence of the pseudo gap in the one-electron density of states due to order-$1/{\cal N}$ self-energy corrections is discussed in Appendix B.

\section{Inverse-degeneracy expansion}
We consider the generalized $\cal N$-orbital Hubbard model:\cite{quantum}
\begin{equation}
H=-t\sum_{\langle ij \rangle, \sigma,\alpha} (a_{i\sigma
\alpha}^{\dagger} a_{j \sigma \alpha} + {\rm H.c.}) + \frac{1}{\cal
N} \sum_{i, \alpha, \beta} (U_1 a^{\dagger}_{i \uparrow \alpha} a_{i
\uparrow \alpha} a^{\dagger}_{i \downarrow \beta} a_{i \downarrow
\beta} + U_2 a^{\dagger}_{i \uparrow \alpha} a_{i \uparrow \beta}
a^{\dagger}_{i \downarrow \beta} a_{i \uparrow \alpha}) \; ,
\end{equation}
where $\alpha,\beta$ refer to the degenerate orbital indices and the factor $1/\cal N$ is included to render the energy density finite in the ${\cal N} \rightarrow \infty$ limit. In the isotropic limit ($U_1=U_2=U$), the two interaction terms (density-density and exchange-type with respect to orbital indices) are together equal to
$U(-{\bf S}_i . {\bf S}_i + n_i ^2)$ in terms of the total spin ${\bf S}_i \equiv \sum_\alpha \psi_{i\alpha}^\dagger
({\mbox{\boldmath $\sigma$}}/2) \psi_{i\alpha}$ and charge $n_i \equiv \sum_\alpha \psi_{i\alpha}^\dagger ({\bf 1}/2)
\psi_{i\alpha}$ operators, and the Hamiltonian is therefore explicitly spin-rotationally symmetric.

With $z$ as the spin-quantization direction, it is convenient to evaluate the time-ordered transverse spin-fluctuation propagator:
\begin{equation}
\chi ^{-+} ({\bf q},\omega) = i \int dt  \; e^{i\omega (t-t')}
\sum_\beta \sum_j e^{i{\bf q}.({\bf r}_i - {\bf r}_j)}
\langle \Psi_0 | {\rm T}[ S_{i\alpha} ^- (t) S_{j\beta} ^+ (t')] \Psi_0 \rangle
\end{equation}
involving the spin-lowering and spin-raising operators $S^\mp = \psi^\dagger (\sigma^\mp/2) \psi$, where $\psi$ is the electron field operator. The transverse propagators $\chi^{-+}$ and $\chi^{+-}$ yield the $x,y$ components of the magnetic response, which are identical to the $z$ response due to spin-isotropy in the paramagnetic ground state $|\Psi_0 \rangle$. When evaluated in the spontaneously-broken-symmetry state, the transverse spin-fluctuation propagator also describes collective spin-wave and particle-hole Stoner excitations.\cite{tri,triself}

In terms of the exact irreducible propagator $\phi({\bf q},\omega)$, which incorporates all self-energy and vertex corrections, the spin-fluctuation propagator can be generally expressed as:
\begin{equation}
\chi^{-+}({\bf q},\omega) = \frac{\phi({\bf q},\omega)}
{1-U\phi({\bf q},\omega)} .
\end{equation}
The inverse-degeneracy expansion:\cite{quantum}
\begin{equation}
\phi = \phi^{(0)} + \left ( \frac{1}{\cal N} \right ) \phi^{(1)} +
\left (\frac{1}{\cal N} \right )^2 \phi^{(2)} + ...
\end{equation}
systematizes the diagrams for $\phi$ in powers of the expansion parameter $1/{\cal N}$ which, in analogy with $1/S$ for quantum spin systems, plays the role of $\hbar$. This expansion explicitly preserves spin-rotational symmetry and therefore the Goldstone mode in the broken-symmetry state, and has been used recently to evaluate quantum corrections to spin-wave energies and spin stiffness in the antiferromagnetic\cite{quantum} and ferromagnetic\cite{singh_ferro} states of the Hubbard model.

\subsection*{$\cal N \rightarrow \infty$ limit}

In the $\cal N \rightarrow \infty$ limit, only the ``classical'' term $\phi^{(0)}\equiv \chi^0$ survives,
and the ladder series with interaction $U_2$ yields the random phase approximation (RPA):
\begin{equation}
\chi^{-+}_{\rm RPA} ({\bf q},\omega) = \frac{\chi^0({\bf q},\omega)}
{1-U\chi^0({\bf q},\omega)} \; ,
\end{equation}
amounting to a classical description of non-interacting spin-fluctuation modes. 
Here the bare antiparallel-spin particle-hole propagator:
\begin{equation}
\phi^{(0)}({\bf q},\omega) \equiv  \chi^0 ({\bf q},\omega) =
\sum_{\bf k} \left( \frac{1} {\epsilon_{\bf k}^{\uparrow +} -
\epsilon_{\bf k - q}^{\downarrow -} - \omega -i \eta} +
\frac{1} {\epsilon_{\bf k-q}^{\downarrow +} -
\epsilon_{\bf k}^{\uparrow -} + \omega -i \eta} \right) \; ,
\end{equation}
where $\epsilon_{\bf k}^\sigma = \epsilon_{\bf k} + U_1 n/2$ are the Hartree-Fock (HF) band energies in the paramagnetic state, and the superscript $+(-)$ refers to particle (hole) states above (below) the Fermi energy $E_{\rm F}$.
The spin-independent HF band-energy shift $U_1 n/2$, corresponding to the $\cal N \rightarrow \infty$ self energy, can be trivially transformed away by an energy shift, as assumed henceforth.

\begin{figure}
\vspace*{0mm}
\includegraphics[height = 70mm, width = 50mm,angle=-90]{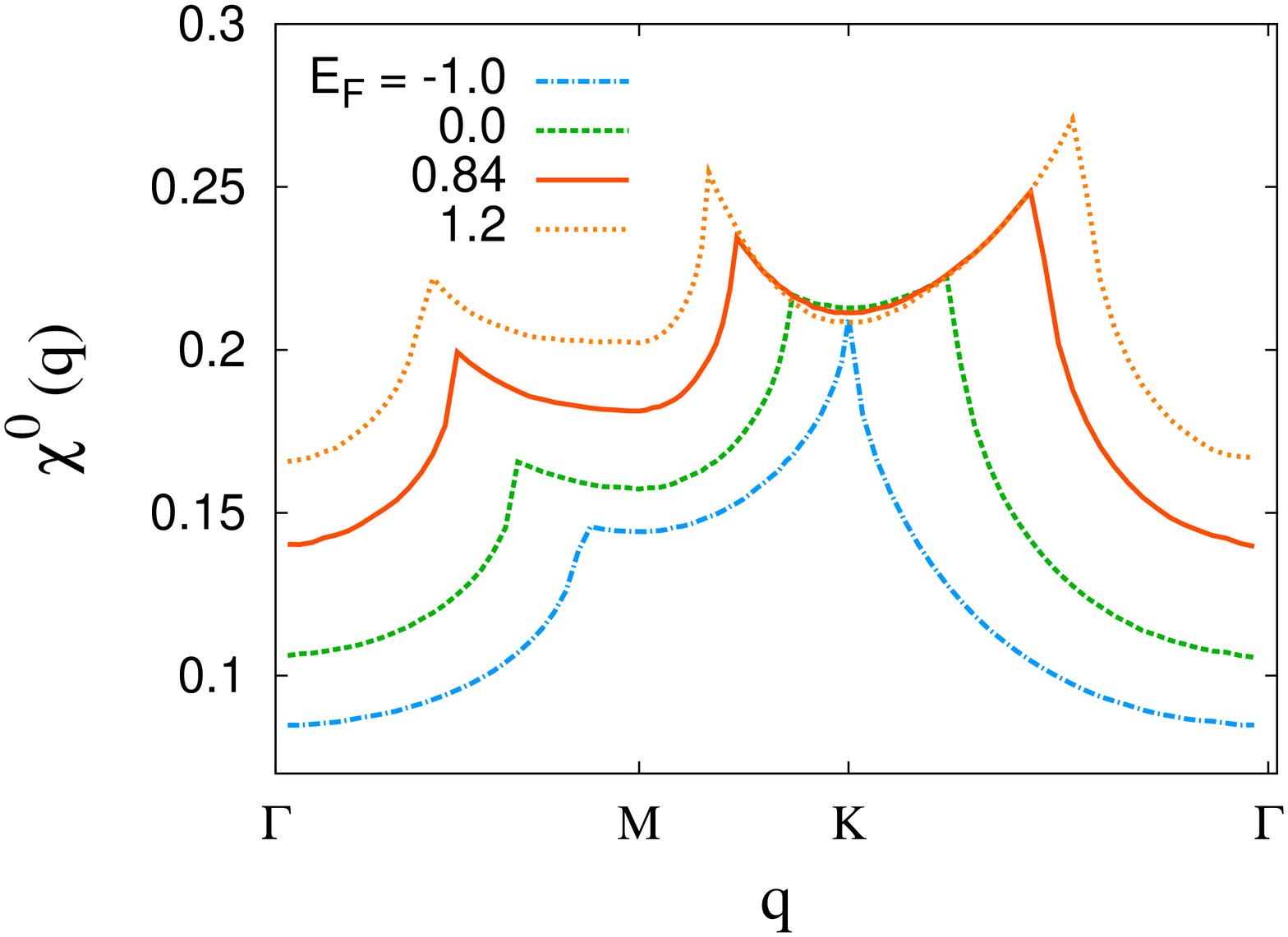}
\includegraphics[height = 50mm, width = 35mm,angle=-90]{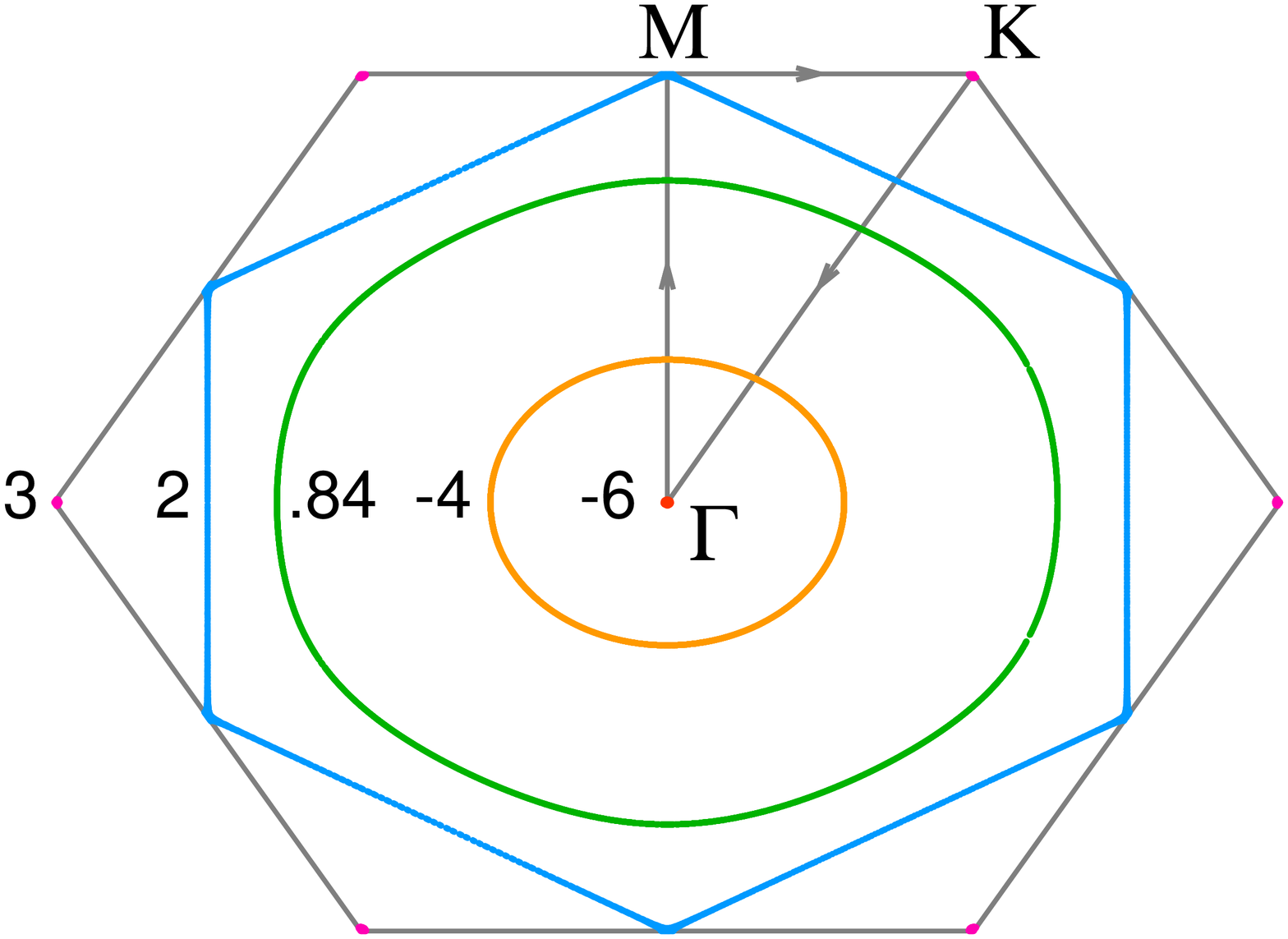}
\caption{The bare magnetic response function at different fillings, showing the relative suppression (enhancement) of the K response (which corresponds to $120^\circ$ AF ordering) with electron (hole) doping. Also shown are few constant-energy contours in the triangular-lattice Brillouin zone.}
\label{chi0dope}
\end{figure}

The bare particle-hole propagator $\chi^0 ({\bf q})$ yields the magnetic response to a static, spatially-varying magnetic field, the rich behaviour of which is shown in Fig. 1 for different Fermi energies, with corresponding fillings $n=0.6,0.8,1.0,1.1$, respectively. Also shown are few constant-energy contours in the triangular-lattice Brillouin zone. 
Contributions to $\chi^0$ from particle states near the nested hexagonal contour ($\epsilon_{\bf k} = 2$)
with divergent density of states is responsible for the cusps in the bare magnetic response function. 
For half filling, the most significant features are the comparable response at the three symmetry points $\Gamma$, M, K, and peaks at the three points approximately midway between them. The magnetic orderings corresponding to the three symmetry points $\Gamma$, M, K are $(0,0,0)$, $(0,\pi,\pi)$, and $(2\pi/3,2\pi/3,2\pi/3)$, respectively, where the triplet corresponds to the ordering wave vector in the three lattice directions. Similarly, for the three midpoints M/2, (M+K)/2, 3K/4, the ordering wave vectors are $(0,\pi/4,\pi/4)$, $(\pi/6,7\pi/6,7\pi/6)$, and $(\pi,\pi/4,\pi/4)$, respectively.

The comparable magnetic response at these six symmetry points, which correpond to very different magnetic orderings, represents the weak-coupling picture of competing orders and magnetic frustration in the triangular-lattice paramagnet.
Note that the response at K, corresponding to $120^\circ$ ordering, is not maximum. In view of the expected instability towards $120^\circ$ AF ordering at strong coupling, it is of particular interest to examine the evolution of the magnetic response with increasing interaction $U$.

\begin{figure}
\vspace*{-20mm} \hspace*{-0mm}
\psfig{figure=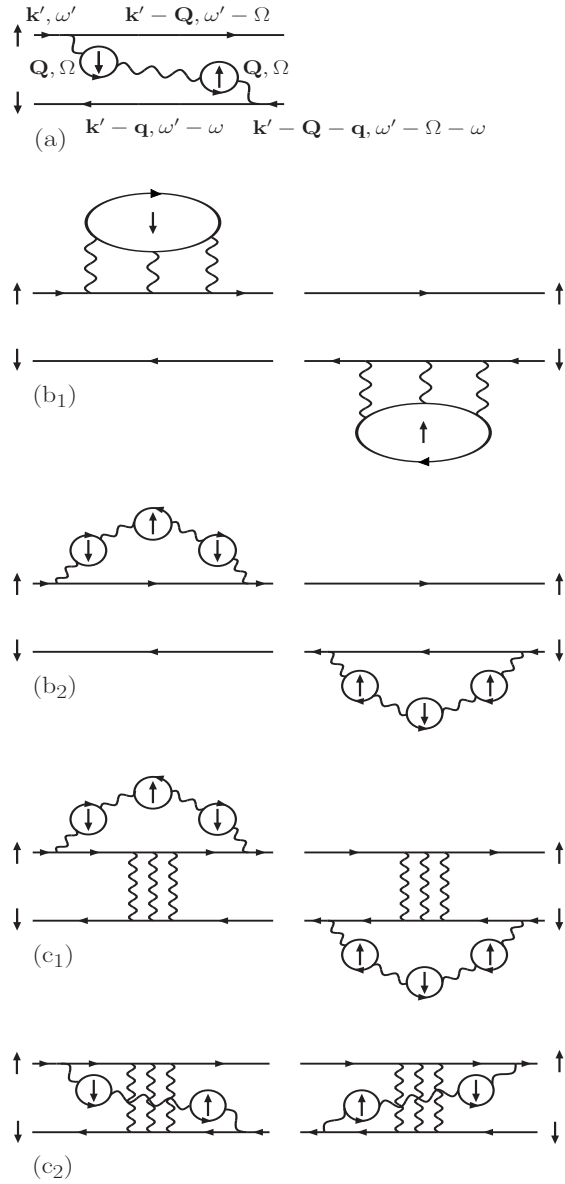,width=160mm} \vspace{-20mm}
\caption{The first-order quantum corrections to the irreducible particle-hole
propagator $\phi({\bf q},\omega)$.}
\label{diagrams}
\end{figure}

\section{$1/{\cal N}$ corrections}

The order-$1/{\cal  N}$ diagrams which yield quantum correction $\phi^{(1)}$ to the irreducible particle-hole propagator $\phi({\bf q},\omega)$ are shown in Figure \ref{diagrams}. The fermion lines represent the HF Green's functions $G^0$. Diagrams $(b_1)$ and $(b_2)$ involve self-energy corrections due to transverse (ladder diagrams) and longitudinal (bubble diagrams) spin fluctuations, respectively. Other self-energy diagrams with $1/{\cal  N}$ corrections to the Hartree self energy, as in the antiferromagnetic state, vanish identically as there are no quantum corrections to particle densities $n_\sigma$ in the paramagnetic state. Diagrams $(a)$ and $(c)$ involve vertex corrections, with (c) representing coupling of longitudinal and transverse spin fluctuations.

The quantum corrections represented by diagrams (a), (b), (c) incorporate different aspects of correlation effects. These include renormalized and dynamical effective interaction (a), negative correction due to spectral-weight transfer and energy renormalization arising from self-energy corrections (b), and negative contribution due to particle-particle correlations of the crossed diagrams (c2). The strong particle-particle correlations found in this study, which suppress the magnetic response through the vertex corrections, are relevant for pairing correlations in the context of the observed superconductivity in the BEDT compounds. 

The corresponding expressions are given below. For diagram $(a)$, we obtain:
\begin{equation}
\phi^{(a)}({\bf q},\omega) = i \int_{-\infty}^{\infty}\frac{d\Omega}{2\pi}
\sum_{\bf Q} \gamma^{(a)}({\bf Q},\Omega) U_{\rm eff}^{\uparrow\downarrow}({\bf Q},\Omega) \; ,
\end{equation}
where the four-fermion vertex:
\begin{equation}
\gamma^{(a)}({\bf Q},\Omega) = i \int \frac{d\omega'}{2\pi}
\sum_{\bf k'} G^0({\bf k'},\omega') G^0({\bf k'-q},\omega'-\omega)
G^0({\bf k'-Q},\omega'-\Omega) G^0({\bf k'-Q-q},\omega'-\Omega-\omega)\; ,
\end{equation}
and the effective antiparallel-spin interaction:
\begin{equation}
U_{\rm eff}^{\uparrow\downarrow}({\bf Q},\Omega)= \frac{U^3 \chi_0^2({\bf Q},\Omega)}{1-U^2 \chi_0^2({\bf Q},\Omega)}
\end{equation}
involves the even-bubble series with interaction $U_1$.

For diagrams $(b_1)$ and $(b_2)$ involving self-energy corrections, we obtain:
\begin{equation}
\phi^{(b)}({\bf q},\omega) = -i \int_{-\infty}^{\infty}\frac{d\Omega}{2\pi}
\sum_{\bf Q} \gamma^{(b)}({\bf Q},\Omega) U_{\rm eff}^{(b)}({\bf Q},\Omega) \; ,
\end{equation}
where the four-fermion vertex:
\begin{equation}
\gamma^{(b)}({\bf Q},\Omega) = i \int_{-\infty}^{\infty}\frac{d\omega'}{2\pi}
\sum_{\bf k'} [G^0({\bf k'},\omega')]^2 G^0({\bf k'-Q},\omega'-\Omega)
[G^0({\bf k'-q},\omega'-\omega)+G^0({\bf k'+q},\omega'+\omega) ] ,\;
\end{equation}
and the effective interaction:
\begin{eqnarray*}
U_{\rm eff}^{(b)} ({\bf Q},\Omega) &=&
\frac{U^2\chi_0({\bf Q},\Omega)}{1-U\chi_0({\bf Q},\Omega)}+
\frac{U^2\chi_0({\bf Q},\Omega)}{1-U^2\chi_0^2({\bf Q},\Omega)} \\
&\equiv& U_{\rm eff}^{-+}({\bf Q},\Omega)+ U_{\rm eff}^{\sigma\sigma}({\bf Q},\Omega)
\end{eqnarray*}
includes the transverse contribution $U_{\rm eff}^{-+}({\bf Q},\Omega)$,
involving the RPA ladder series (with interaction $U_2$) and the parallel-spin contribution
$U_{\rm eff}^{\sigma\sigma}({\bf Q},\Omega)$, involving the RPA odd-bubble series (with interaction $U_1$).

Finally, for the vertex correction diagrams $(c_1)$ and $(c_2)$, involving both the ladder series (with interaction $U_2$) and the bubble series (with interaction $U_1$), we obtain:
\begin{eqnarray}
\phi^{(c1)}({\bf q},\omega) = &-i&
\!\!\!\int_{-\infty}^{\infty}\frac{d\Omega}{2\pi} \sum_{\bf Q}
[ \gamma^{(c)+} ({\bf Q},\Omega) \gamma^{(c)+} ({\bf Q},\Omega) + 
\gamma^{(c)-} ({\bf Q},\Omega) \gamma^{(c)-} ({\bf Q},\Omega) ] \nonumber \\ &\times&
\left(\frac{U}{1-U\chi_0({\bf Q},\Omega)}\right)
\left(\frac{U^2\chi_0({\bf q-Q},\omega-\Omega)} {1-U^2\chi_0^2({\bf
q-Q},\omega-\Omega)}\right) \; , \nonumber \\
\ \\
\phi^{(c2)}({\bf q},\omega) = &+i&
\int_{-\infty}^{\infty}\frac{d\Omega}{2\pi} \sum_{\bf Q}
2 \gamma^{(c)+} ({\bf Q},\Omega) \gamma^{(c)-} ({\bf Q},\Omega) \nonumber \\
&\times& \left(\frac{U}{1-U\chi_0({\bf
Q},\Omega)}\right)\left(\frac{U} {1-U^2\chi_0^2({\bf
q-Q},\omega-\Omega)}\right),
\end{eqnarray}
where the three-fermion vertices:
\begin{equation}
\gamma^{(c)\pm} ({\bf Q},\Omega) = i
\int_{-\infty}^{\infty}\frac{d\omega'}{2\pi} \sum_{\bf k'}
G^0({\bf k' \pm q},\omega' \pm \omega) G^0({\bf k'},\omega')
G^0({\bf k' \pm Q},\omega' \pm \Omega) \; .
\end{equation}

It is straightforward to show, using transformations such as $({\bf k'},\omega') \rightarrow ({\bf k'+q},\omega'+\omega)$
and ${\bf k',Q} \rightarrow -{\bf k'},-{\bf Q}$, that the quantum corrections $\phi^{(a),(b),(c)}$ are symmetric in the ${\bf q},\omega$ space:
\begin{equation}
\phi({\bf -q},-\omega) = \phi({\bf q},\omega) = \phi({\bf -q},\omega) .
\end{equation}

\begin{figure}
\vspace*{0mm} \hspace*{0mm}
\includegraphics[height = .3\columnwidth,angle=-90]{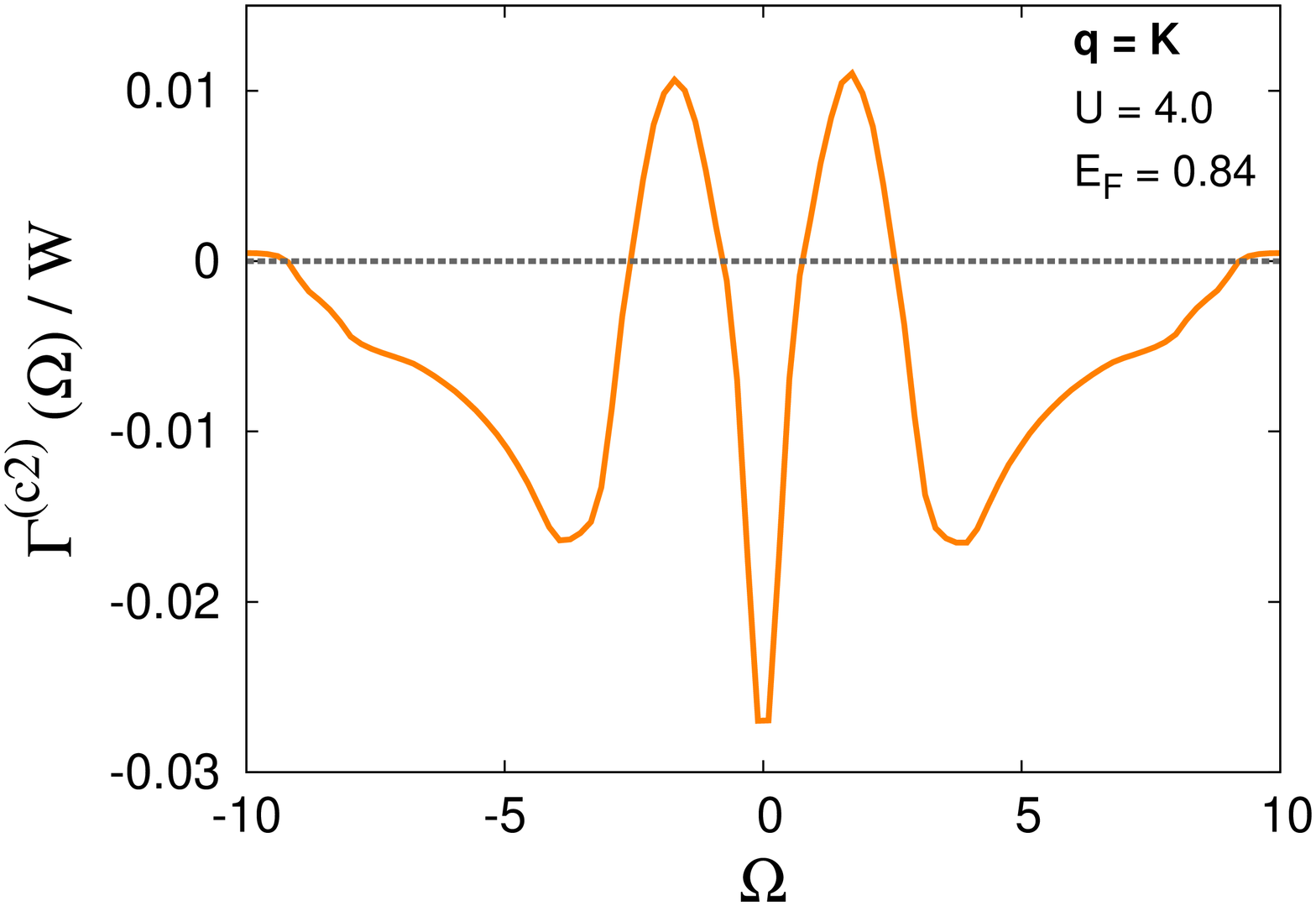}
\vspace*{0mm} \hspace*{0mm}
\includegraphics[height = .3\columnwidth,angle=-90]{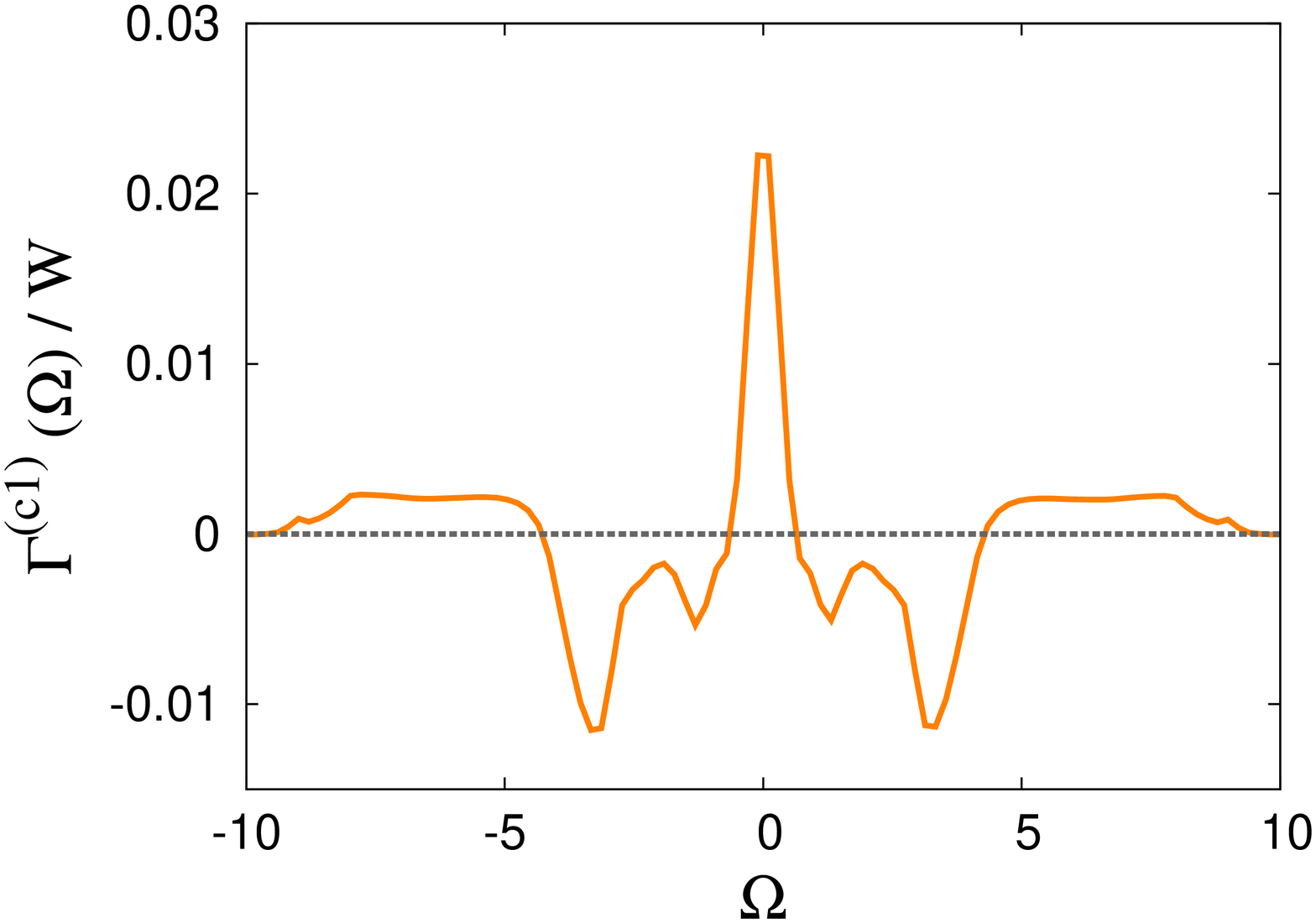}
\vspace*{0mm} \hspace*{0mm}
\includegraphics[height = .3\columnwidth,angle=-90]{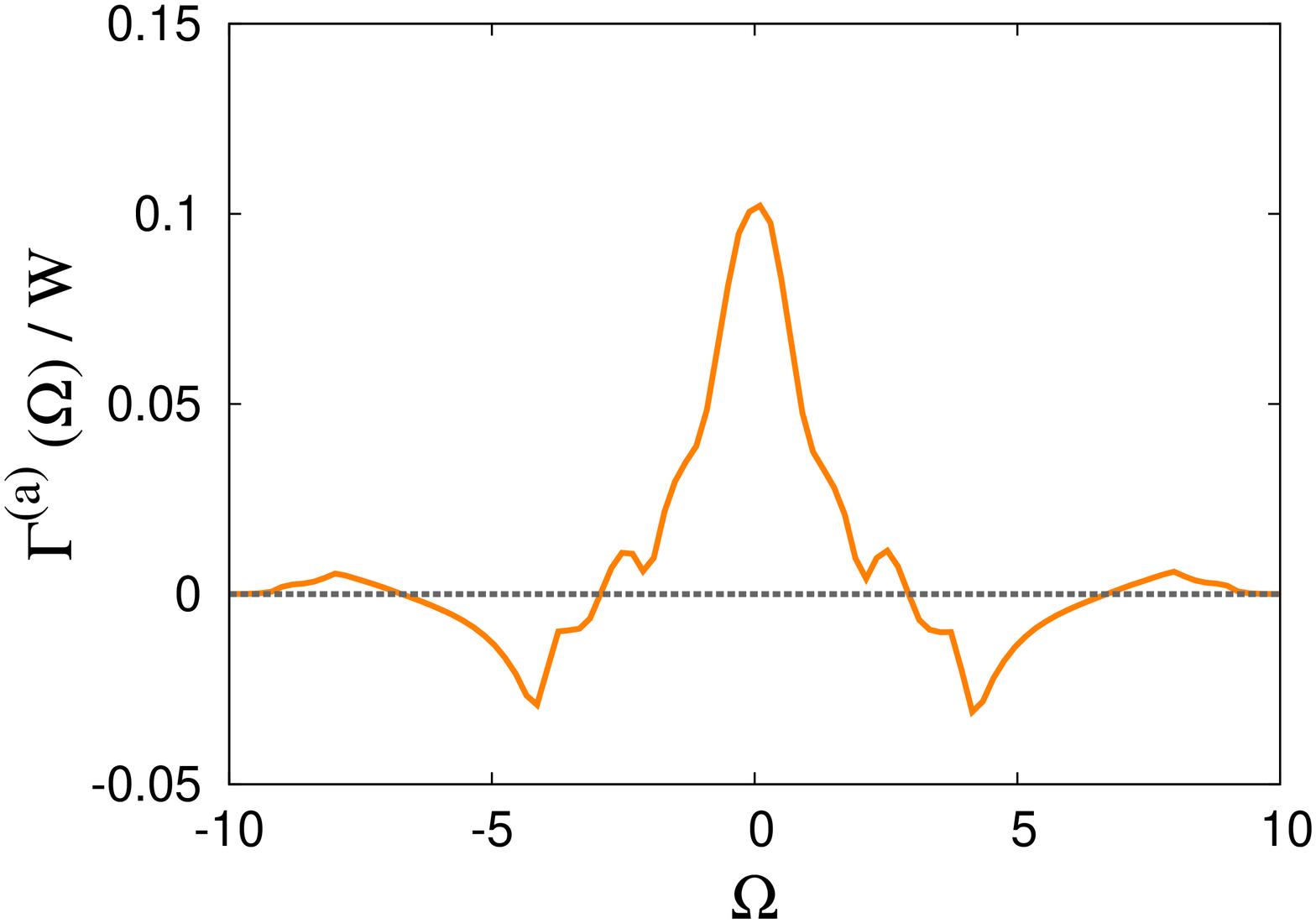}
\caption{The $\Omega$-resolved contributions of $\phi^{(c2)}$, $\phi^{(c1)}$, and $\phi^{(a)}$, showing enhanced contributions at low frequency (paramagnon enhancement) as well as at intermediate and high frequencies.} \label{symm}
\end{figure}

\section{Results at half filling}

In this section we present results at half filling for the order-$1/{\cal N}$ contributions $\phi^{(a),(b),(c)}$ in the static limit and discuss the evolution of the static magnetic response with increasing interaction strength $U$. Evaluation of the fermion energy-momentum integrals for the three- and four-fermion vertices (Eqs. 8,11,14), corresponding to integrating out the fermion degrees of freedom, is illustrated in Appendix A. In our numerical calculations for $\phi^{(a),(b),(c)}$ we have taken grid sizes $dk' = dQ = 0.1$, $d\Omega=0.2$, and $\eta=0.1$, on an energy scale $t=1$. 

In order to examine the relative contributions to the three quantum corrections $\phi^{(a),(b),(c)}$ from the different internal bosonic modes (involving ladders and bubbles) and the three- and four-fermion vertices $\gamma^{(c)}$ and $\gamma^{(a),(b)}$, we introduce functions $\Gamma^{(a),(b),(c)}(\Omega)$ defined by:
\begin{equation}
\phi^{(a),(b),(c)} = \frac{1}{W} \int_{-\infty} ^{\infty} d\Omega \;
\Gamma^{(a),(b),(c)}(\Omega) ,
\end{equation}
where $W$ is the fermion bandwidth. The functions $\Gamma(\Omega)$ effectively yield combined density of states of the internal excitations involving the vertex functions and the bosonic modes. Typical plots for $\Gamma(\Omega)/W$ are shown in Fig. \ref{symm}. The symmetric-$\Omega$ behaviour provides a numerical check for the calculations. It also shows that the intermediate- and high-$\Omega$ contributions are quite significant and comparable to the low-$\Omega$ contributions which show the usual sharp paramagnon enhancement.

\begin{figure}
\vspace*{0mm}
\hspace*{0mm}
\includegraphics[width = 0.25\columnwidth,angle=-90]{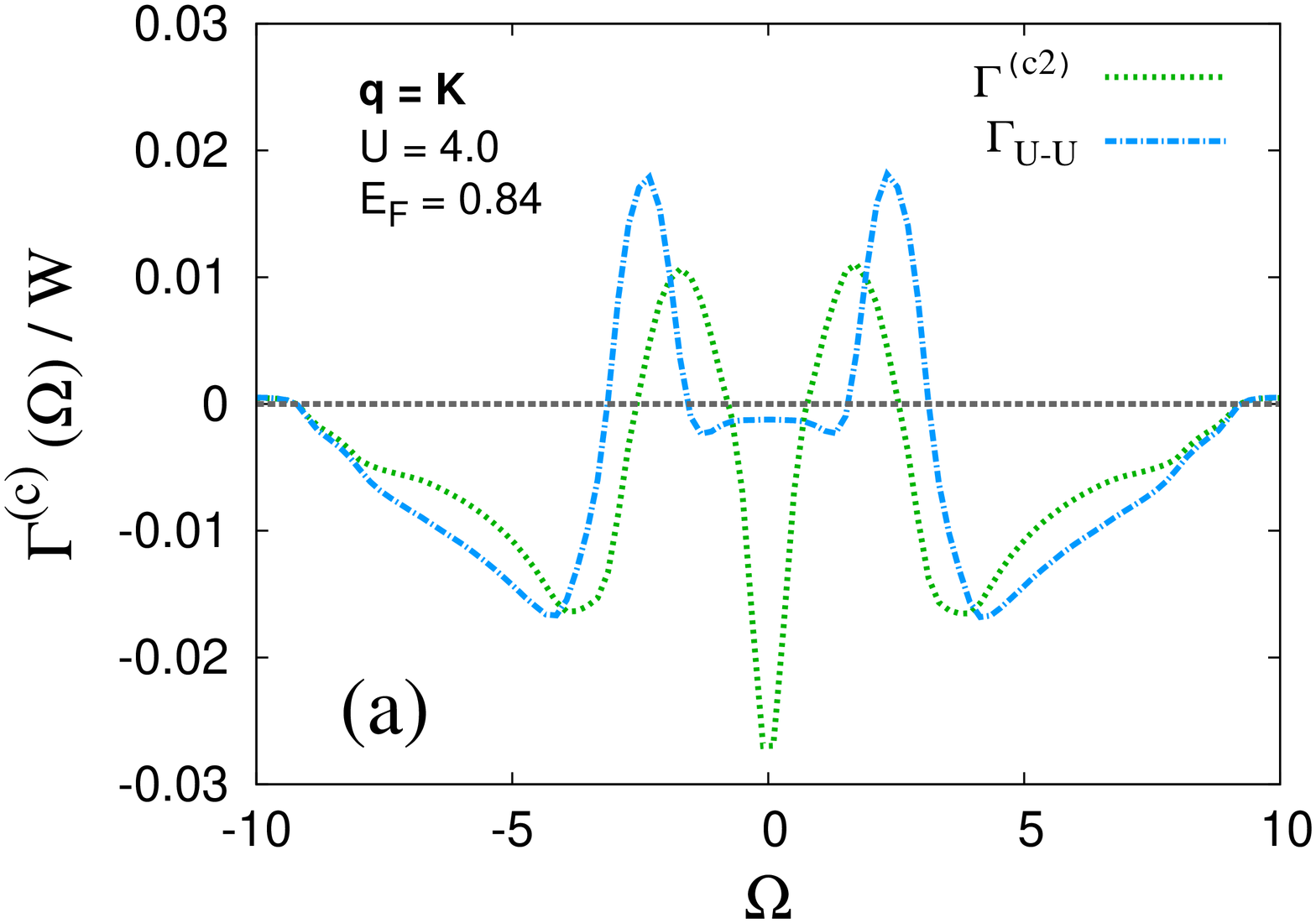}
\vspace*{0mm}
\hspace*{0mm}
\includegraphics[width = 0.25\columnwidth,angle=-90]{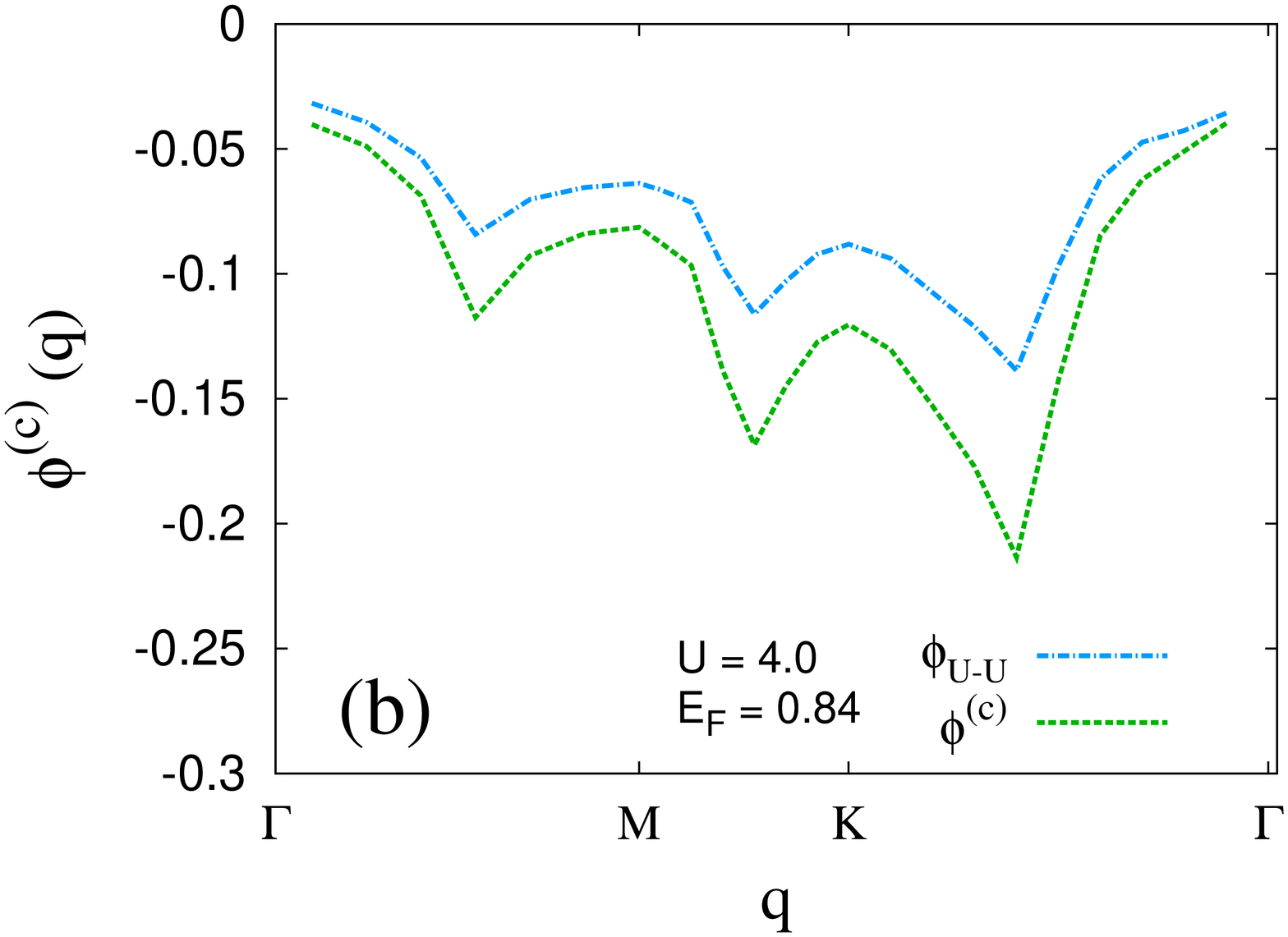}
\caption{Comparison of $\phi^{(c)}$ with the second-order result, with respect to the
$\Omega$-resolved contribution (a) and the $\mathbf{q}$ dependence (b).} \label{gc12u}
\end{figure}

The $\Omega$-resolved contribution of $\phi^{(c)}$ with only the leading, second-order $(U^2$) term in Eq. (13) is also shown in Fig. \ref{gc12u}(a) for comparison. Apart from the missing low-$\Omega$ paramagnon enhancement as expected,
it strongly resembles the $\Omega$-structure of the full $\phi^{(c2)}$, implying an essentially fermionic origin for the intermediate- and high-energy structures. Also shown (b) is a comparison of the $\mathbf{q}$ dependence of $\phi^{(c)}$ with the second-order result. The exactly same $\mathbf{q}$-structure implies that the fermionic terms $\gamma^{(c)}$ are fully responsible for the characteristic momentum dependence as well. Renormalization of the internal paramagnon mode, within a self-consistent approach, is therefore not expected to qualitatively change this $\mathbf{q}$-structure.

\begin{figure}
\vspace*{-10mm}
\hspace*{-0mm}
\epsfig{figure=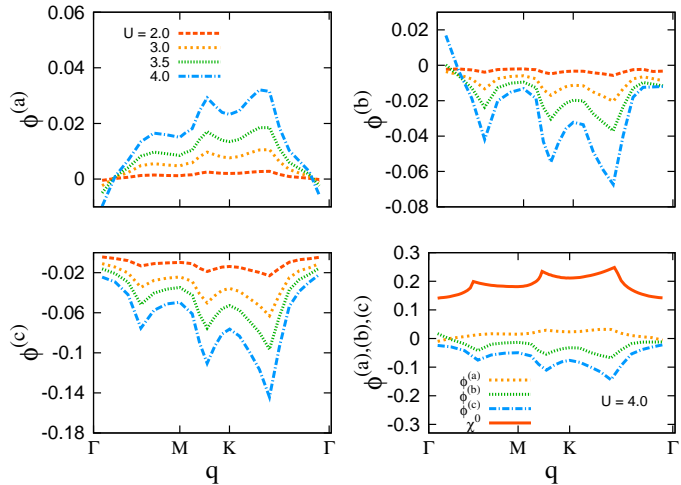,width=80mm,angle=-90}
\vspace*{-0mm}
\hspace*{0mm}
\caption{Evolution of the different contributions $\phi^{(a)}$, $\phi^{(b)}$, and $\phi^{(c)}$ with $U$, showing enhanced negative contribution of $\phi^{(b)}$ and $\phi^{(c)}$ at the same $\bf q$ points where $\chi^0({\bf q})$ peaks.}
\label{phiabc}
\end{figure}

Evolution of the different contributions $\phi^{(a)}$, $\phi^{(b)}$, and $\phi^{(c)}$ with increasing interaction strength $U$ is shown in Fig. \ref{phiabc}, along with their relative comparison for $U=4$. An enhanced negative contribution of $\phi^{(b)}$ and $\phi^{(c)}$ is seen to occur at the same $\bf q$ points where $\chi^0({\bf q})$ peaks. It should also be noted that although the self-energy contribution (b) has similar momentum dependence, the largest contribution comes from the vertex correction (c). Furthermore, the relative difference between the K and peak contributions are significantly enhanced as compared to that in $\chi^0$. In order to make contact with the single-orbital Hubbard model, double counting at the $U^2$ level in diagrams (c2) and (b) was  removed in the Fig. 5 calculations.

\begin{figure}
\epsfig{figure=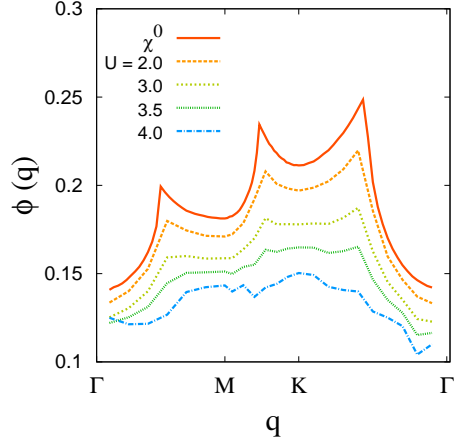,width=60mm,angle=-90} \caption{Evolution of the magnetic response function $\phi({\bf q})$ with $U$, showing the inversion of the curvature around K and M with increasing $U$, with maximum response at K indicating stabilization of the $120^{\circ}$-ordered AF state in the strong coupling limit.}
\label{phiq}
\end{figure}

In order to examine the $U$-evolution of the full static magnetic response, we evaluate the irreducible propagator $\phi$ to all orders within an approximate resummation scheme:\cite{ma}
\begin{equation}
\phi ({\bf q}) = \frac{\chi^0({\bf q})}
{1 - \phi^{(1)}({\bf q})/\chi^0({\bf q})}
\end{equation}
which is exact to first order in the $1/{\cal N}$ expansion.
Fig. \ref{phiq}, which summarizes the main result of this paper,
shows the evolution of $\phi ({\bf q})$ with increasing $U$. The enhanced negative contribution
of the quantum corrections $\phi^{(b)}$ and $\phi^{(c)}$ --- at the same $\bf q$ points where the bare response $\chi^0({\bf q})$ peaks --- results in an inversion of the curvature around K and M with increasing $U$. The net response is maximum at K, indicating stabilization of the $120^{\circ}$-ordered AF state in the strong-coupling limit. This is consistent with the consensus of a $120^\circ$-ordered AF ground state for the equivalent $S=1/2$, nearest-neighbour quantum Heisenberg antiferromagnet (QHAF) on a triangular lattice.\cite{order1,loop1,order2,loop2,qmc}

The change in the curvature of the response function with increasing $U$ implies that it goes through a regime of nearly flat magnetic response around K, as seen in Fig. \ref{phiq}. This is in agreement with the observed lack of dispersion in recent PIRG calculations, where the NMI state has been suggested to be a new type of degenerate quantum spin phase having gapless and dispersionless (flat) spin excitations, indicating a high momentum degeneracy and accounting for the quantum melting of simple translational symmetry breakings including the AF long-ranged order.\cite{pirg3} 

This evolution of the magnetic response function highlights a complex feature of correlation effects on magnetic frustration in the triangular-lattice Hubbard model. Initially, increasing interaction strength results in enhanced competing interactions and magnetic frustration, as indicated by the decreasing (negative) curvature of the magnetic response,
which even becomes flat in a broad momentum range around K, with the degenerate response indicating high degree of spin disorder. However, beyond a critical interaction strength, the magnetic response develops increasingly positive curvature around K, indicating build-up of $120^\circ$ AF spin correlations and suppression of magnetic frustration. In the strong-coupling limit, long-range $120^\circ$ AF order sets in, as only the geometrical frustration of the triangular lattice remains due to the surviving NN AF spin couplings in the equivalent QHAF. 

The overall suppression of the magnetic response with increasing $U$ is a manifestation of correlation effects, arising mainly from the particle-particle correlations involved in the vertex correction (c2) with crossed interaction lines and, to a relatively smaller extent, also from the pseudo-gap formation due to self-energy corrections (b). From the paramagnetic side, the leading instability towards the $120^\circ$ AF state at strong coupling and at half filling is thus a consequence of these vertex and self-energy corrections.

The maximum magnetic response at K implies onset of AF spin correlations with short-range $120^\circ$ ordering. What is the effect of these correlations on the self-energy correction and the electronic density of states? Using characteristic band dispersion identities for the triangular-lattice, an approximate analytical estimate for the electron self energy due to spin-fluctuation scattering yields a two-band structure, with similar dispersion as for the broken-symmetry state, and band separation increasing with interaction strength, eventually leading to the insulating gap.\cite{pseudogap}

As the renormalized magnetic response is maximum at K, from Eq. (3) we can estimate the critical interaction strength 
$U^* = 1/\phi(K) \approx 1/0.14 \approx 7$ for the magnetic transition to the $120^\circ$ ordered AF state. In this estimation we have assumed that the renormalized {\em internal} bosonic excitations have the simple form $\chi({\bf Q},\Omega) = \frac{\chi^0({\bf Q},\Omega)} {1-U'\chi^0({\bf Q},\Omega)}$, as considered within the two-particle self-consistent (TPSC) approximation,\cite{tpsc} with the 
renormalized interaction $U' \gtrsim 4$ corresponding to the maximum bare response $\chi^0 \lesssim 0.25$. This estimate for $U^*$ is in good agreement with the value obtained earlier ($\gtrsim 6$) from the broken-symmetry side by considering 
the melting of magnetic order due to quantum spin fluctuations in the $120^\circ$ ordered AF state.\cite{tri}


\section{Finite doping}

In the preceeding section, the renormalized magnetic response at half-filling was shown to be maximum at K, in accord with the consensus that $120^\circ$ AF ordering is stabilized in the strong-coupling limit. In this section, we will examine the effects of finite hole and electron doping on the magnetic response function, and therefore on the stability of the $120^\circ$ ordered AF state. Earlier studies have shown that the $120^\circ$ ordered AF state is stabilized for hole doping and the spin stiffness is enhanced, whereas it is destabilized for any amount of electron doping.\cite{tri} These studies were carried out in the broken-symmetry state, with doped holes/electrons introduced in the AF bands within a rigid-band approximation, and effects of finite doping on transverse spin fluctuations were examined by including the {\em intraband} particle-hole processes in the particle-hole propagator.

\begin{figure}
\begin{center}
\vspace*{0mm}
\includegraphics[height = 70mm, width = 50mm,angle=-90]{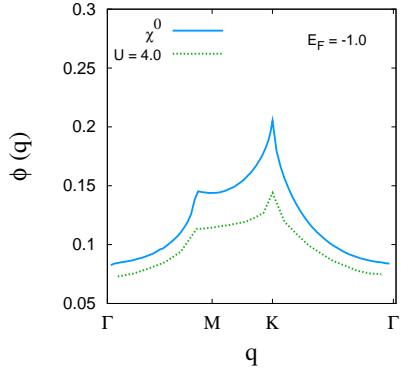}
\end{center}
\caption{The renormalized magnetic response function $\phi({\bf q})$ for $U = 4.0$ and hole doping ($E_{\rm F} = -1.0$)
along with the bare response function $\chi^{0}({\bf q})$, showing that the maximum response remains at K, corresponding to $120^{\circ}$ AF ordering.} \label{hole_doped}
\end{figure}

Fig. \ref{hole_doped} shows a comparison of the bare and renormalized magnetic response for $E_{\rm F}=-1.0$ (hole doping) and $U=4$. The bare magnetic response shows a pronounced peak at  K corresponding to $120^\circ$ AF ordering, indicating drastic suppression of frustration. We find that since quantum corrections are significantly suppressed, this feature survives at the renormalized level, indicating that the dominant magnetic instability at K remains unchanged.

On the other hand, for electron doping we find a subtle inversion of the magnetic response near K on a {\em small-momentum scale}, as shown in Fig. 8(a), indicating destabilization with respect to long-wavelength fluctuations {\em about} the $120^\circ$ ordering, in agreement with earlier results.\cite{tri} We emphasize that although the bare magnetic response also shows a negative curvature around K, indicating relative instability of the $120^\circ$ ordered state, the quantum correction introduces a negative curvature on a much smaller momentum scale. Fig. 8(b) shows that this small-momentum feature arises from the self energy term in $\phi^{(b)}$, and is completely absent in the vertex correction $\phi^{(c)}$.

\begin{figure}
\includegraphics[width = 0.30\columnwidth,angle=-90]{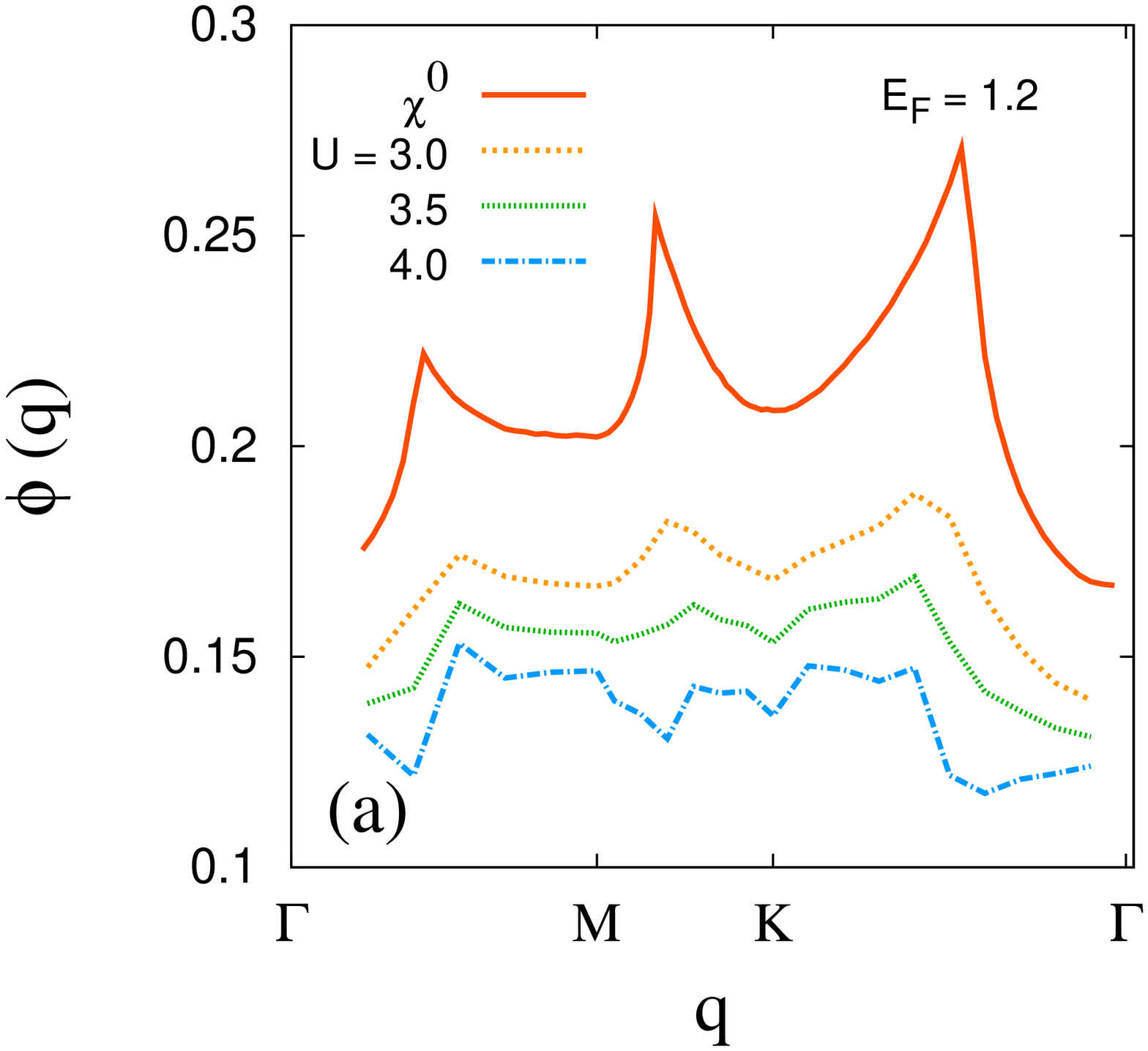}
\includegraphics[width = 0.30\columnwidth,angle=-90]{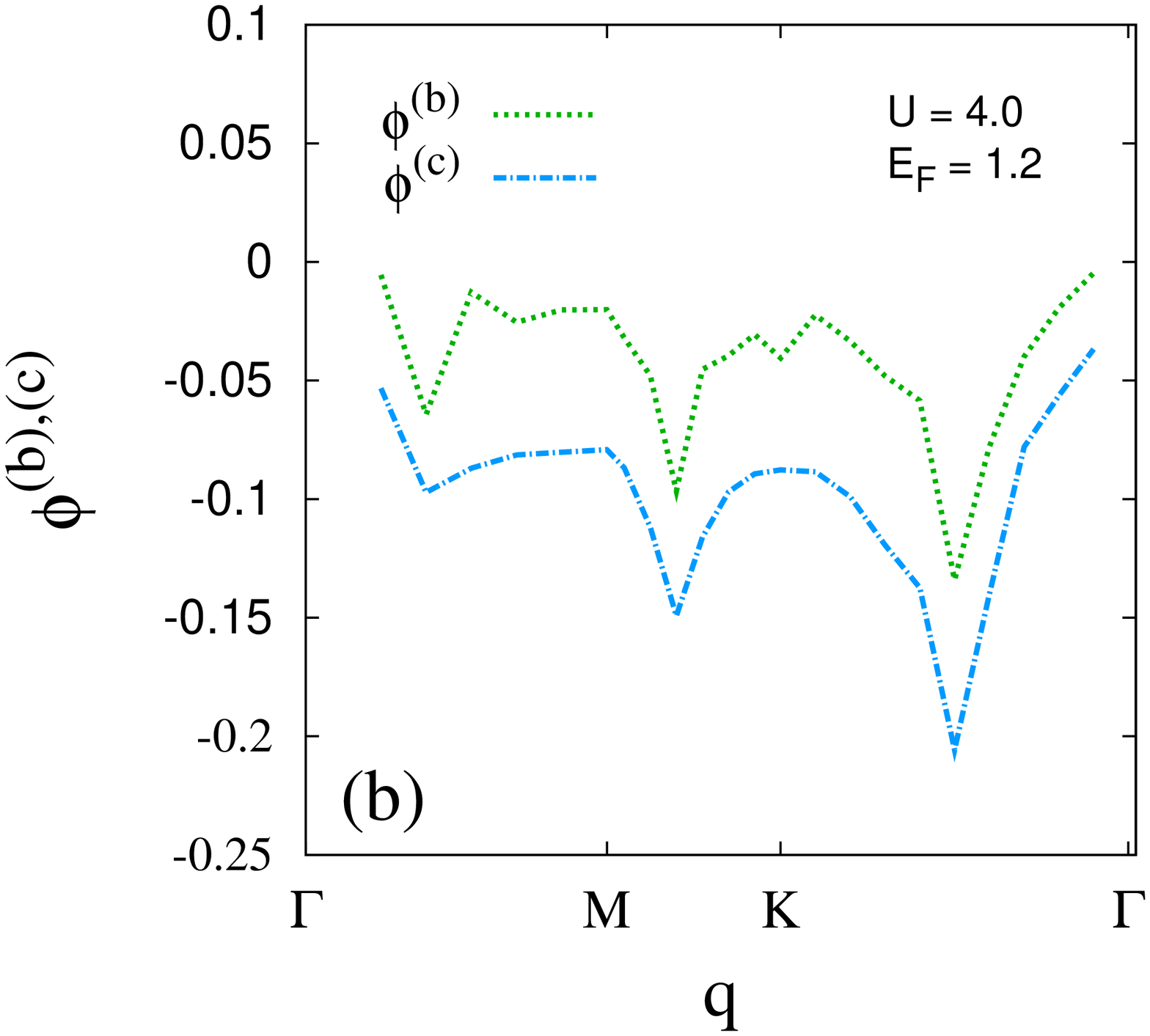}
\caption{Evolution of $\phi({\bf q})$ with $U$ for the electron-doped case shows a dip near K, 
indicating destabilization of the $120^{\circ}$ ordered state on a small-momentum scale (a);
this small-momentum feature near K originates from the self-energy contribution $\phi^{(b)}$ and not from 
the vertex correction $\phi^{(c)}$ (b).}
\label{phiqeldp}
\end{figure}



\newpage
\section{Conclusions}
In conclusion, we have investigated correlation effects on magnetic frustration in the triangular-lattice Hubbard model by studying the evolution of the magnetic response function with increasing interaction strength. We have employed a systematic inverse-degeneracy expansion scheme which preserves spin-rotational symmetry and therefore allows seamless interpolation into the broken-symmetry state.

The strong suppression of the renormalized magnetic response, mainly by vertex corrections due to particle-particle correlations and to a smaller extent by self-energy corrections due to pseudo-gap formation, was shown to result in an inversion of the magnetic response function with respect to the bare result, yielding a maximum response at the K point, consistent with the expected $120^\circ$ AF instability at half filling and strong coupling.
In view of the comparable and locally maximum response at M, it is interesting to note that 
a $\pi$-flux spin-liquid state, which on spinon condensation leads to ordering wave vector on the Brillouin zone edge centers (M points), has been proposed for the $J-J'$ QHAF on the triangular lattice.\cite{wang_2006}

Hole doping was shown to suppress competing interaction and frustration effects, resulting in enhanced stability for $120^\circ$ AF ordering. On the other hand, for electron doping, a small-momentum feature was obtained in the magnetic response function near K  corresponding to instability of the $120^\circ$ ordering with respect to long-wavelength fluctuations. These results at half filling and finite doping are in agreement with earlier results obtained in the broken-symmetry AF state, indicating that this 
spin-rotationally-symmetric approach can be used reliably and seamlessly between the paramagnetic and broken-symmetry states.

A self-consistent analysis for both the irreducible particle-hole propagator and the self energy, incorporating $120^\circ$ AF spin correlations in the renormalized internal bosonic modes $({\bf Q},\Omega)$, will highlight the effects of these correlations on dynamical features such as the spin-fluctuation energy scale (especially for the important low-energy modes), the spectral function ${\rm Im} \chi^{-+}(\bf{q},\omega)$ of magnetic excitations (the integrated weight of which yields the local spin correlations $\langle S^- S^+ \rangle$ and local moments in the paramagnetic state), as well as the spin-fluctuation self energy and pseudo-gap formation in the electronic density of states. A finite-frequency study dealing with these dynamical aspects is currently in progress. 

\appendix
\section{evaluation of the 3-fermion vertex}
We illustrate here the evaluation of the fermion terms by integrating out the fermion energy-momentum degrees of freedom for the three-fermion vertex. Including all possible retarded/advanced cases, we obtain:
\begin{eqnarray}
\gamma^{(c)+} ({\bf Q},\Omega) &=& i\int_{-\infty}^{\infty}\frac{d\omega'}{2\pi} \sum_{\bf k'}
G^0({\bf k'+q},\omega'+\omega) G^0({\bf k'},\omega')
G^0({\bf k'+Q},\omega'+\Omega) \nonumber \\
&=& i\int_{-\infty}^{\infty}\frac{d\omega'}{2\pi} \sum_{\bf k'}
\frac{1}{\omega'+\omega-\epsilon_{\bf k'+q}^\pm \pm i\eta} .
\frac{1}{\omega'-\epsilon_{\bf k'}^\pm \pm i\eta} .
\frac{1}{\omega'+\Omega-\epsilon_{\bf k'+Q}^\pm \pm i\eta} \nonumber \\
&=& i^2 \sum_{\bf k'}
\frac{1}{\epsilon_{\bf k'+q}^- - \epsilon_{\bf k'}^+ -\omega + i\eta} .
\frac{1}{\epsilon_{\bf k'+q}^- - \epsilon_{\bf k'+Q}^+ + \Omega -\omega + i\eta}
\nonumber \\
&+& i^2 \sum_{\bf k'}
\frac{1}{\epsilon_{\bf k'}^- - \epsilon_{\bf k'+q}^+ +\omega + i\eta} .
\frac{1}{\epsilon_{\bf k'}^- - \epsilon_{\bf k'+Q}^+ + \Omega + i\eta}
\nonumber \\
&+& i^2 \sum_{\bf k'}
\frac{1}{\epsilon_{\bf k'+Q}^- - \epsilon_{\bf k'+q}^+ - \Omega + \omega + i\eta} .
\frac{1}{\epsilon_{\bf k'+Q}^- - \epsilon_{\bf k'}^+ - \Omega + i\eta}
\nonumber \\
&-& i^2 \sum_{\bf k'}
\frac{1}{\epsilon_{\bf k'+q}^+ - \epsilon_{\bf k'}^- -\omega - i\eta} .
\frac{1}{\epsilon_{\bf k'+q}^+ - \epsilon_{\bf k'+Q}^- + \Omega -\omega - i\eta}
\nonumber \\
&-& i^2 \sum_{\bf k'}
\frac{1}{\epsilon_{\bf k'}^+ - \epsilon_{\bf k'+q}^- +\omega - i\eta} .
\frac{1}{\epsilon_{\bf k'}^+ - \epsilon_{\bf k'+Q}^- + \Omega - i\eta}
\nonumber \\
&-& i^2 \sum_{\bf k'}
\frac{1}{\epsilon_{\bf k'+Q}^+ - \epsilon_{\bf k'+q}^- - \Omega + \omega - i\eta} .
\frac{1}{\epsilon_{\bf k'+Q}^+ - \epsilon_{\bf k'}^- - \Omega - i\eta} \; .
\end{eqnarray}
The ${\bf k'}$ summations were performed numerically over the triangular-lattice Brillouin zone with a grid size $dk' = 0.1$.

\section{O$(1/{\cal N})$ Self-energy correction}
In section IV it was mentioned that the negative contribution of $\phi^{(b)}$ arises from the redistribution of spectral weight due to self-energy corrections. We illustrate here this feature in the renormalized density of states resulting from the first-order $(1/{\cal N})$ self-energy correction:
\begin{equation}
\Sigma_{\bf k}(\omega) =
U^2 \sum_{\bf Q} \int \frac{d\Omega}{2\pi}
\left [\frac{\chi_0({\bf Q},\Omega)}{1-U\chi_0({\bf Q},\Omega)} +
\frac{\chi_0({\bf Q},\Omega)}{1-U^2\chi_0 ^2({\bf Q},\Omega)} \right ]
G^0 ({\bf k-Q},\omega-\Omega) 
\end{equation}
due to exchange of transverse and longitudinal spin fluctuations, corresponding to ladder and bubble diagrams, respectively. The two (retarded and advanced) contributions to the self energy,
\begin{equation}
\Sigma^{R}_{\bf k}(\omega) =
U^2 \sum_{\bf Q} \int_0 ^\infty \frac{d\Omega}{\pi}
{\rm Im} [\chi_{\rm total}({\bf Q},\Omega)]_{\rm R}
\frac{1}{\omega-\Omega-\epsilon_{\bf k-Q}^+ +i\eta }
\end{equation}
and
\begin{equation}
\Sigma^{A}_{\bf k}(\omega) =
U^2 \sum_{\bf Q} \int_{-\infty} ^0 \frac{d\Omega}{\pi}
{\rm Im} [\chi_{\rm total}({\bf Q},\Omega)]_{\rm A}
\frac{1}{\omega+|\Omega|-\epsilon_{\bf k-Q}^- -i\eta } ,
\end{equation}
correspond to the intermediate fermion state ${\bf k-Q}$ lying inside ($-$) or outside ($+$) the Fermi surface, and the total spin-fluctuation term $[\chi_{\rm total}({\bf Q},\Omega)]$ includes both the ladder and bubble contributions.
The retarded self energy $\Sigma^{R}_{\bf k}(\omega)$ yields negative imaginary part only for $\omega > E_{\rm F}$, whereas
the advanced self energy $\Sigma^{A}_{\bf k}(\omega)$ yields positive imaginary part only for $\omega < E_{\rm F}$.

\begin{figure}
\epsfig{figure=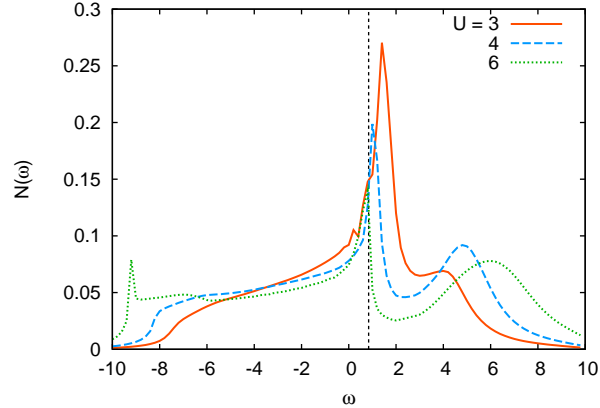,width=80mm}
\caption{Evolution of the renormalized density of states evaluated from Eq. (B5), showing the opening of the pseudo gap with increasing interaction strength $U$.}
\label{rendos}
\end{figure}

The total self energy $\Sigma_{\bf k}(\omega) = \Sigma^{R}_{\bf k}(\omega) +  \Sigma^{A}_{\bf k}(\omega)$ yields the renormalized Green's function:
\begin{equation}
G_{\bf k}(\omega) = \frac{1}{\omega - \epsilon_{\bf k} - \Sigma_{\bf k}(\omega)}
\end{equation}
and the one-particle density of states:
\begin{equation}
N(\omega) = \frac{1}{\pi} \sum_{\bf k} \frac{ {\rm Im} \Sigma_{\bf k}(\omega) }
{[\omega-\epsilon_{\bf k} -{\rm Re} \Sigma_{\bf k}(\omega) ]^2 + [{\rm Im} \Sigma_{\bf k}(\omega) ]^2 } \; .
\end{equation}

If the intermediate fermion states ${\bf k-Q}$ predominantly lie outside (inside) the Fermi surface for ${\bf k}$ inside (outside), as is characteristic of the unfrustrated square lattice for ${\bf Q}$ near the AF ordering wavevector $(\pi,\pi)$, then the negative (positive) contribution of $\Sigma^{R(A)}_{\bf k}(\omega)$ pulls down (pushes up) the hole (electron) states in energy, resulting in the opening of an energy gap in the one-particle density of states. Fig. \ref{rendos} shows the emergence of a pseudo gap in the renormalized one-particle density of states with increasing interaction strength $U$.

\end{document}